\documentclass[11pt]{article}
\usepackage{graphicx}
\usepackage{caption}
\usepackage[letterpaper,margin=20mm]{geometry}
\usepackage{amsmath}
\usepackage[inline]{enumitem}
\usepackage{rotating}
\usepackage{hyperref}
\usepackage{natbib}
\usepackage{subcaption}
\newcommand{\footremember}[2]{%
    \footnote{#2}
    \newcounter{#1}
    \setcounter{#1}{\value{footnote}}%
}

\title{A speciation simulation that partly passes open-endedness tests}
\author{
  Théo de Pinho\footremember{depinho}{theo.depinho.etu@univ-lille.fr - University of Lille, France}
  \and Lana Sinapayen\footremember{sinapayen}{lana.sinapayen@gmail.com - Sony Computer Science Laboratories / National Institute for Basic Biology, Japan}
  }

\begin{document}

\maketitle

\textbf{Abstract.} One of the main goals of artificial life research is to recreate in artificial systems the trends for ever more complex and novel entities, interactions and processes that we see in Earth's biosphere, that is, to create \textit{open-ended} systems. In this paper, we test for Tokyo type 1 open-ended evolution (OEE) of the Tree of Life Simulation (ToLSim), an artificial life software created by Lana Sinapayen. To do so, we conducted an experiment to measure evolutionary activity statistics. These require us to define the notion of components. Here, we define components as the agent's ``genes''. The results show that ToLSim is capable of exhibiting unbounded total cumulative evolutionary activity. However, total and median \textit{normalized} cumulative evolutionary activity appear bounded and new evolutionary activity is persistently null, suggesting that ToLSim is not open-ended. Further studies on ToLSim could repeat this experiment with individuals or even species, rather than genes, to test whether the present results are valid.

\section{Introduction}
\label{section/introduction}

Artificial life (ALife) is a field that seeks to understand life and its fundamental properties \citep{bedau-2007-artificial}. One of its main subject of study is \textit{open-ended evolution} (OEE), specifically in life and life-like systems. OEE denotes evolutionary systems that continually renew and complexify themselves \citep{bedau-2000-open, stanley-2017-last}. Earth's biosphere is an example of (probable) open-ended system. From simple, microscopic beings to complex animal societies, natural selection of Earth life seems to produce more and more complex entities and systems, without ever running out of novel ways to do so. Culture also exhibits some form of open-endedness \citep{borg-2024-evolved, chalmers-2010-high, buchanan-2011-measuring, oka-2015-open}. One might wonder if OEE can be recreated from scratch in artificial systems. Doing so could not only improve our understanding of open-endedness's necessary and sufficient conditions, but also allows us to create open-ended algorithms that generate never-thought-of-before ideas in arts and science\footnote{We do not endorse innovation as an end. Innovations as an end can have harmful results. But we think that innovation can, in some circumstances, increase the quality of life and the happiness of sentient beings by making their life less painful and, potentially, more interesting.}\citep{stanley-2019-open}. Attempts at creating open-ended artificial systems have been made in chemistry (``wet" ALife), robotics (``hard" ALife) and computer science (``soft" ALife) since several decades \citep{bedau-2007-artificial}, but no artificial life system seems to have achieved the level of open-endedness of Earth's biosphere \citep{karelin-2025-fermi}. In computer science, specifically, signs of open-endedness have been demonstrated in only a small number of programs, like Geb \citep{channon-2003-improving, channon-2006-unbounded}.

In this paper, we test the open-endedness of the Tree of Life Simulation (ToLSim), a multi-agent program created by Sinapayen that simulates the emergence of speciation \footnote{The source code for the simulation and analysis can be found at \url{https://github.com/LanaSina/speciation}. A video of a run of ToLSim can be found at \url{https://www.youtube.com/watch?v=GJsBIk-9iyE}}. This is done by applying steps 1, 2, and 3 of Alastair Channon's procedure for testing for Tokyo type 1 OEE \citep{channon-2024-procedure}, Tokyo type 1 OEE being characterized by ``interesting new kinds of entities and interactions'' \citep{packard-2019-overview}.

The procedure implies the use of evolutionary activity statistics. Indeed, evolutionary activity statistics appear to be quite a good proxy of open-endedness. They measure core aspects of open-endedness: novelty, complexity, and accumulation of adaptive success. Furthermore, evolutionary activity statistics seem to be both sensible \textit{and} specific. Indeed, they confirm that the biosphere is open-ended and that many artificial systems (e.g. Bugs and Evita) are not \citep{bedau-1997-comparison}.

Evolutionary activity statistics measure the evolutionary activity of components. Components can be genes, individuals, or species, depending on the level of analysis. Some evolutionary activity statistics are the \textit{total cumulative evolutionary activity}, the \textit{mean cumulative evolutionary activity}, the \textit{diversity} and the \textit{new activity}. These will be detailed in section (\ref{section/methods}).

One of the first use of evolutionary activity statistics can be found in \citet{bedau-1992-measurement} where the authors used them to study a soft ALife system of \textit{strategic bugs}.
\citet{bedau-1997-comparison} used evolutionary activity statistics to study three systems: two artificial life programs (Evita and Bugs), and the biosphere (indicated by the fossil record). They measured \textit{diversity}, \textit{total cumulative activity} and \textit{mean cumulative activity}. The biosphere showed strong increasing trends for total cumulative activity and diversity, but the Evita and Bugs artificial systems did not. 
\citet{bedau-1998-classification} then added a new statistic, \textit{new evolutionary activity}, which provides a measure of novelty of evolutionary systems. They also added a test that uses three classes of evolutionary systems, which they defined as follows: evolutionary systems that show no adaptive evolutionary activity (class 1), ones that show bounded adaptive evolutionary activity (class 2) and ones that show unbounded adaptive evolutionary activity (class 3). The statistics could thus be used to show whether a system had unbounded evolutionary trends (i.e. whether it were open-ended).
Later, \citet{channon-2006-unbounded} used evolutionary activity statistics to test the Geb artificial life program and found that it showed open-ended trends.

In the remainder of this paper, we apply evolutionary activity statistics to test ToLSim's open-endedness.
In the next section, we explain how ToLSim works.
In the Methods section, we show how we apply evolutionary activity statistics to test it.
Finally, in the Results section, we present the key results of our experiment and draw conclusions on ToLSim's open-endedness.

\section{The Tree of Life Simulation (ToLSim)}
\label{section/tolsim}

\begin{figure}[!htbp]
    \centering
    \includegraphics[width=10cm]{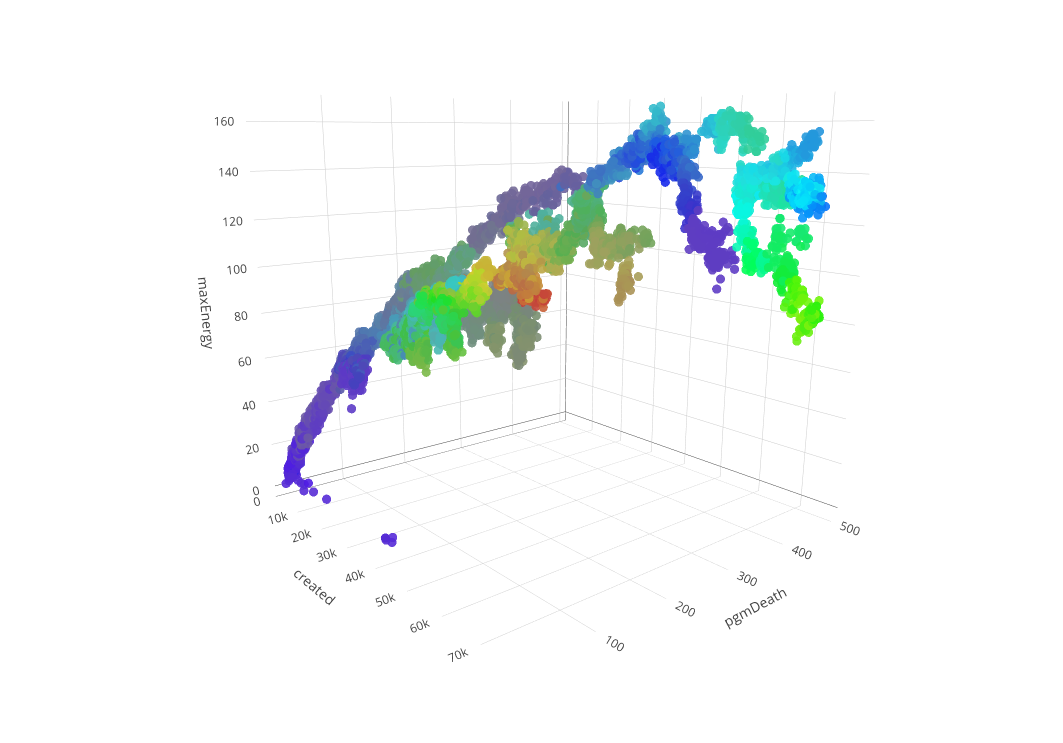}
    \caption{A tree representing a run of ToLSim. Each dot represents an individual. Individuals are plotted according to the moment of their birth (\texttt{created}), their maximum energy (\texttt{maxEnergy}) and their maximum or ``programmed'' death (\texttt{pgmDeath}). Branches are visible, highlighting the emergence of different ``species''.}
    \label{fig/tolsim-tree}
\end{figure}

The Tree of Life Simulation (ToLSim) is an artificial life software that simulates the emergence of different branches of species. The statistical plot in figure \ref{fig/tolsim-tree} shows what a ``tree of life'' can look like in ToLSim. The program is written in Java, with R code that serves to analyze the produced data.

\begin{figure}[!htbp]
    \centering
    \begin{subfigure}{0.45\textwidth}
        \centering
        \includegraphics[width=\linewidth]{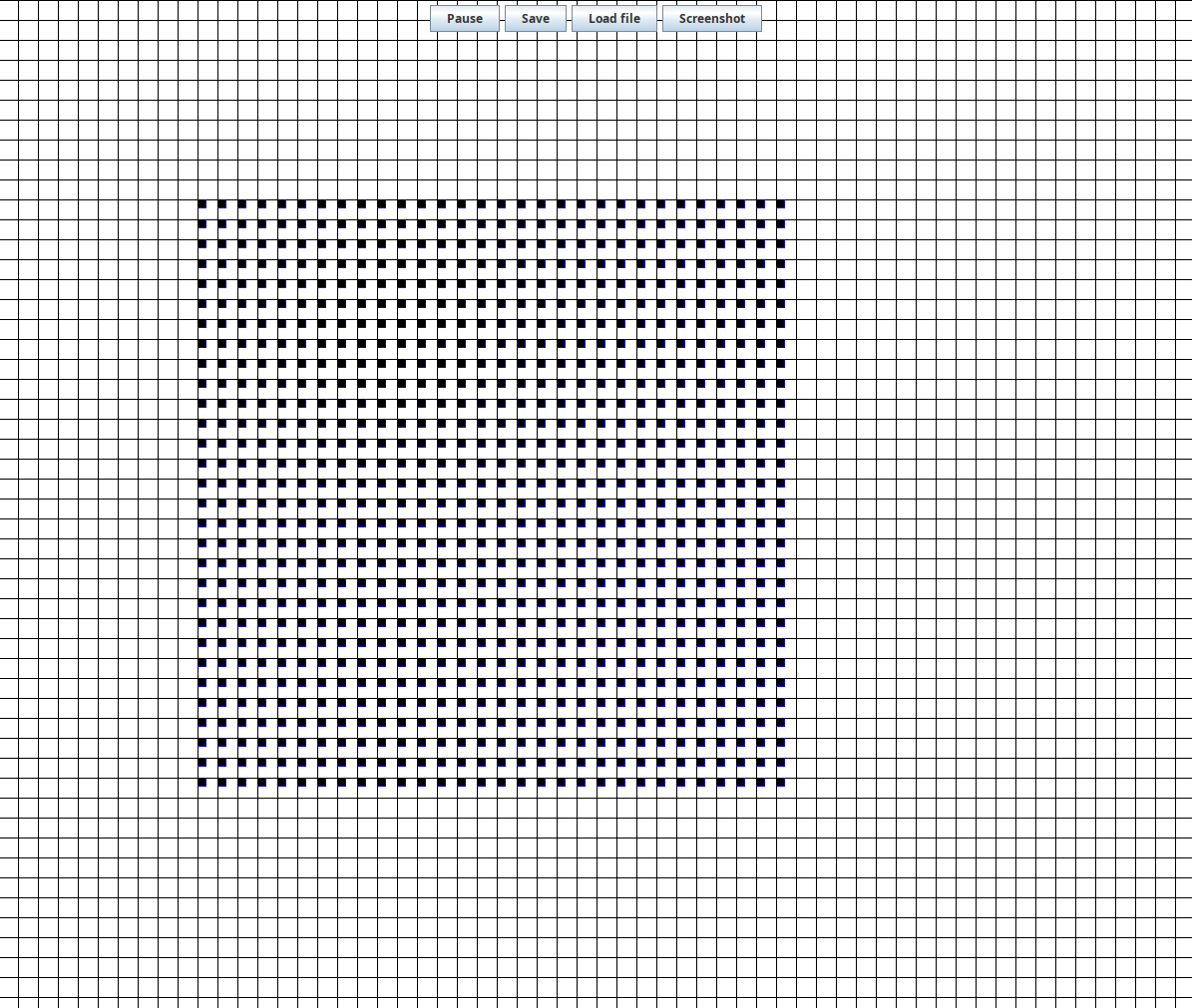}
        \caption{}
    \end{subfigure}\hfill
    \begin{subfigure}{0.45\textwidth}
        \centering
        \includegraphics[width=\linewidth]{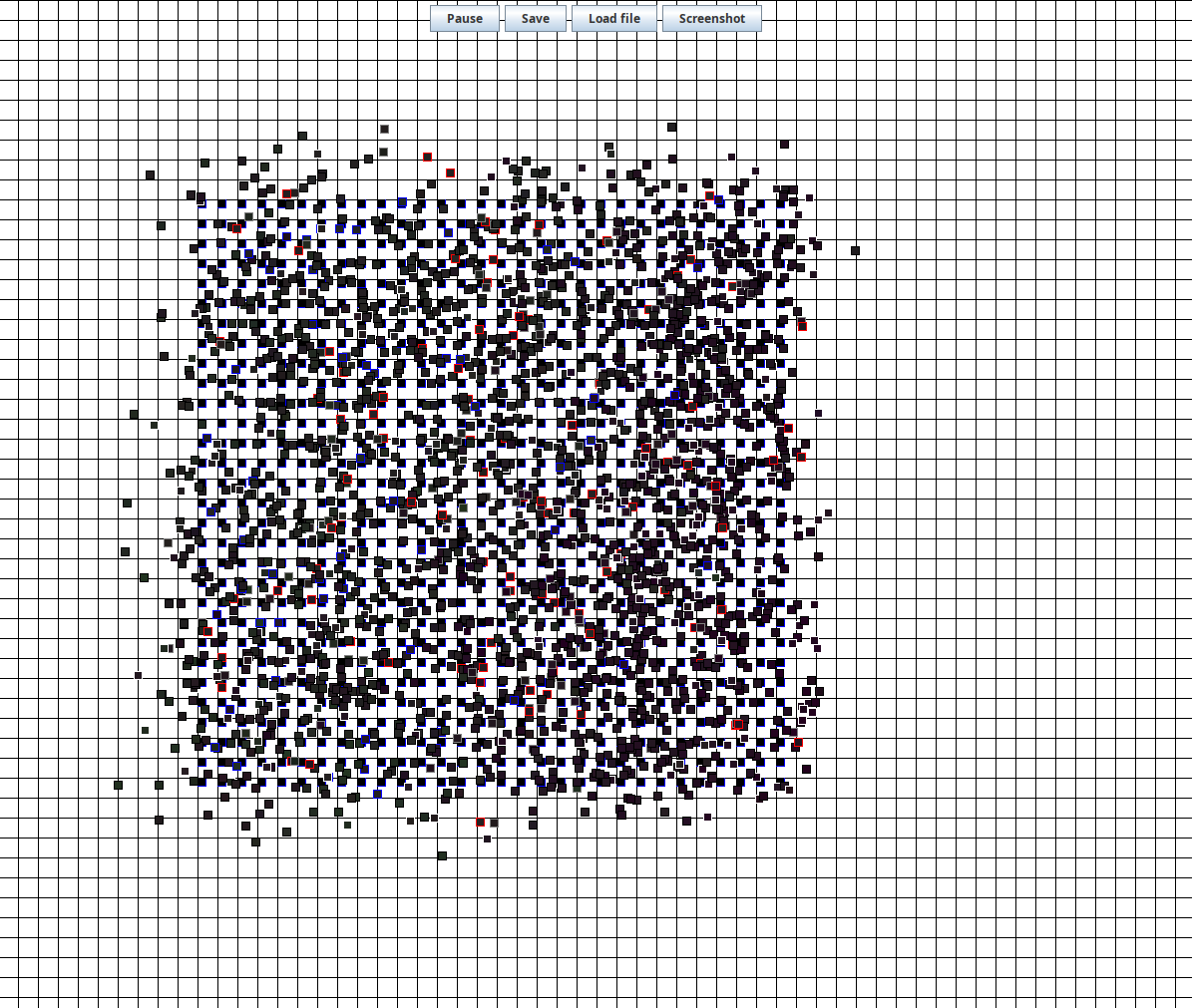}
        \caption{}
    \end{subfigure}

    \medskip 

    \begin{subfigure}{0.45\textwidth}
        \centering
        \includegraphics[width=\linewidth]{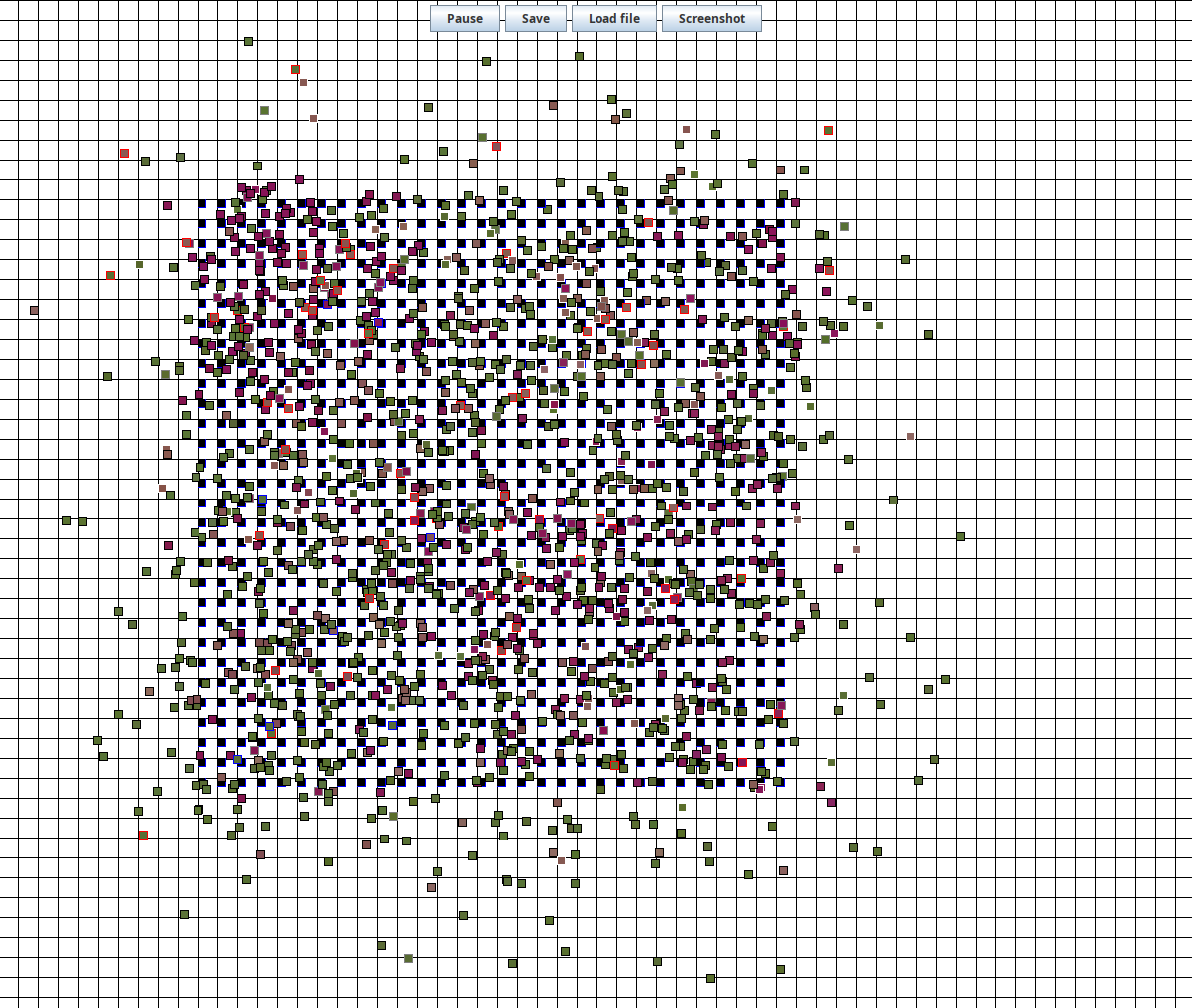}
        \caption{}
    \end{subfigure}\hfill
    \begin{subfigure}{0.45\textwidth}
        \centering
        \includegraphics[width=\linewidth]{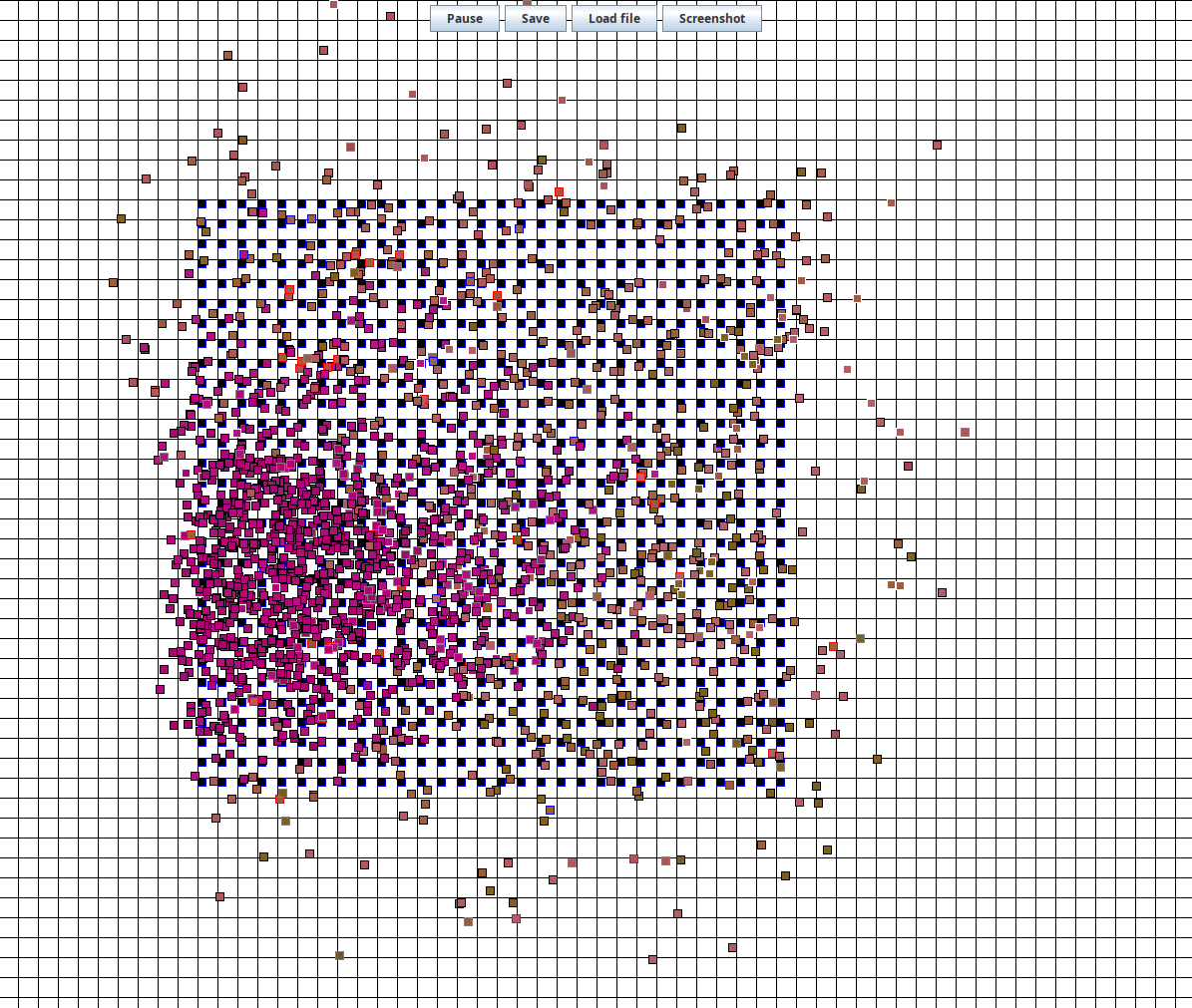}
        \caption{}
    \end{subfigure}

    \caption{Screenshots of a run of ToLSim at different times. (a) shows ToLSim before the apparition of ALife (the black squares are the sources of energy). (b) shows ToLSim just after ALife has appeared and spread (the additional squares are agents). (c) and (d) shows ToLSim several minutes after that. (d) shows a species spreading.}
    \label{fig/tolsim-screenshots}
\end{figure}

The simulation consists of a grid of 50 by 50 cells. At the center of this map, a 30 by 30 area continuously produces agents with random characteristics. At the start of the simulation, there is no evolution; instead, random agents appear spontaneously in the central 30 by 30 area. These can move, try to eat other agents, reproduce by cloning, and die. At first, most agents die without reproducing because they are not well adapted. However, sometimes a first-generation individual succeeds in reproducing (i.e., at cloning itself), thus producing second-generation individuals. These can themselves either die without offspring or produce children. After some time (usually a few hundred time steps from the start of the run, i.e. only a few seconds), an endless chain of generations starts forming itself. This moment is analogous to the apparition of life on Earth, since it is the instant when naturally selected individuals emerge (previous individuals were random). The user interface that comes with the program shows the grid and allows one to clearly witness this instant. Indeed, because first-generation individuals are intentionally made invisible on the grid, the grid is initially lifeless and then, in a matter of seconds, becomes full of moving creatures (figure \ref{fig/tolsim-screenshots}). From there, the population reaches a plateau and continues to evolve.

Let us go into more details about the mechanics of ToLSim. Agents have hereditary properties and non-hereditary properties. The hereditary properties, which are fixed -- meaning they do not vary during the life of an agent -- are the following:
\begin{itemize}
    \item \texttt{speed}: how much the agent can move in the grid at each time step;
    \item \texttt{maxEnergy}: the maximum amount of energy that the agent can store;
    \item \texttt{kidEnergy}: the energy that the agent will spend to produce one offspring;
    \item \texttt{nKids}: the maximum number of kids that the agent can have at each time step;
    \item \texttt{pgmDeath}: the maximum or ``programmed'' age (in time steps) that the agent will be able to reach;
    \item \texttt{matForKids}: the age (in time steps) at which the agent reaches maturity, that is, the age at which it can reproduce;
    \item \texttt{sensors}: sensors of various types used to decide whether to eat another agent or not.
\end{itemize}
These properties are transmitted to offspring. Each time an agent clones itself -- which it does at each time step where it has enough energy to do so --, it has a 50\% chance of producing a mutant offspring. Each trait of a mutant offspring has a 60\% chance of being modified from the value inherited from its parent\footnote{This means that the actual probability of a mutant offspring is less than 50\%: there is a chance that no individual property was randomly mutated.}.

The only non-hereditary property is \texttt{energy}, which is simply the amount of energy an agent has at a given time. An agent gains energy by eating other agents. It loses energy when it moves, when it tries to eat another agent that has more energy than itself, when another agent tries to eat it (if the predator agent succeeds, the prey loses all of its energy and dies), and when it produces offspring. In addition, an agent loses energy at every step simply by maintaining its existence. Finally, an agent dies either when its energy level reaches 0 or when it reaches its maximum lifespan (\texttt{pgmDeath}).

Let us detail how predation works. An agent can only eat agents that are in the same cell as itself. Furthermore, the more agents are in a cell, the more likely the predation between two given agents becomes. In order to attempt to eat another agent, an agent (which, from here, we will call a \textit{predator}) must first detect it, which it can do with its sensors. Each agent has a list of sensors. Each sensor can sense one property of the prey (e.g. \texttt{speed}, \texttt{maxEnergy}, etc.) if that property's value is contained within a range given by:

\begin{align}
    predatorSensorValue-5 \leq preyProperty \leq predatorSensorValue+5
    \label{eq-sensors}
\end{align}

An agent may have several sensors dedicated to the same property, but covering different ranges.

Another agent is detected if at least one of the predator's sensors detects it. For instance, if the predator has a \texttt{speed} sensor with the value 10, it is able to detect agents with a speed of 5, 15 or 9, but it cannot detect agents with a speed of 4, 16 or 50 (except if other sensors detect them). If at that time step the predator has more energy than the prey, the predation attempt succeeds, and the predator absorbs all of the prey's energy. If predation fails, both agents are ``wounded'', that is, they lose energy (the amount of energy each agent loses is equal to the amount of energy of the other agent multiplied by some constant). Note that what an agent can sense does not depend on the agent's own internal properties. Finally, agents are greedy: they will attempt to eat all agents that they detect on their cell.

\section{Methods}
\label{section/methods}

\citep{packard-2019-overview} defined the Tokyo types of OEE as the following:
\begin{enumerate*}[label=(\arabic*)]
    \item interesting new kinds of entities and interactions;
    \item evolution of evolvability;
    \item major transitions and
    \item semantic evolution.
\end{enumerate*}
\citep{channon-2024-procedure} proposed a procedure for testing for OEE Type 1. In this section, we apply the first three steps of this procedure to test ToLSim's open-endedness.

The evolutionary activity statistics used in the procedure require a definition of components. We choose to define components as genes, rather than agents or species. In ToLSim, what best fits the definition of genes are the heritable properties described in section \ref{section/tolsim}: \texttt{speed}, \texttt{maxEnergy}, \texttt{kidEnergy}, \texttt{nKids}, \texttt{death}, \texttt{matForKids}. So, one gene is one property value (e.g. \texttt{speed = 3}, \texttt{nKids = 6}, etc.). Note that because the values of the hereditary properties do not have bounds besides having to be positive, each one of the properties can theoretically correspond to an infinite number of genes.

\subsection{Step 1: basic evolutionary activity statistics}

Step 1 of Channon's procedure involves computing the total cumulative evolutionary activity, which ``provides a measure of the extent of evolutionary activity : how much activity has accumulated and remained in the system through the persistent use of components'' \citep{channon-2006-unbounded}. It is computed in the following way. At each time step $t$, each component $i$ is associated to an \textit{activity increment} $\Delta_i(t)$, which simply reflects the existence or non-existence of the component at time $t$:

\begin{align}
    \Delta_i (t) =
    \left\{
        \begin{array}{rl}
            1 & \text{if component } i \text{ exists at } t \\
            0 & \text{otherwise}
        \end{array}
    \right.
    \label{eq-delta}
\end{align}

Then, the \textit{cumulative evolutionary activity} $a_i(t)$ reflects the persistence of component $i$ over time:

\begin{align}
    a_i (t) =
    \left\{
        \begin{array}{ll}
            \sum_{\tau = 0}^{t} \Delta_i (\tau) & \text{if component } i \text{ exists at } t \\
            0 & \text{otherwise}
        \end{array}
    \right.
    \label{eq-a}
\end{align}

This variable is null if component $i$ does not exist at time $t$; otherwise, it is equal to the total number of time steps during which the component has existed so far. Finally, the sum of cumulative evolutionary activities of all components gives us the \textit{total cumulative evolutionary activity}:

\begin{align}
    A_{cum} (t) = \sum_{i} a_i (t)
    \label{eq-A_cum}
\end{align}

The test for step 1 involves testing whether the total cumulative evolutionary activity is bounded or unbounded. If it is unbounded, then the test for step 1 is passed, and we can continue to step 2.

\subsection{Steps 2 and 3: component-normalized evolutionary activity statistics}

Step 2 of the procedure proposed by Channon involves implementing a neutral ``shadow model'' and a shadow-resetting method. A neutral shadow model (or simply a shadow model) is a model that is executed in parallel to the real model and mimics it: each birth in the real model causes a birth in the shadow model and each death in the real model causes a death in the shadow model. The difference between the two models is that each birth/death in the shadow model is operated on a random component (each component having the same probability to be chosen). Thus, each component that survives in the shadow model does not survive thanks to an adaptive advantage, but rather due to chance. Since the shadow model does not implement natural selection, comparing the shadow model with the real model allows us to evaluate the part of evolutionary activity that is genuinely adaptive.

At regular intervals (e.g. every $1000$ time steps), a shadow-resetting method resets the state (i.e. the population) and the evolutionary activity history of the shadow's components to those of the real model's components. This procedure allows for the shadow model not to derive too far away from the real model and to remain relatively representative of the part of the real evolutionary activity that isn't adaptive.

The shadow model is used to normalize the evolutionary activity of components, i.e. to exclude non-adaptive evolutionary activity. By subtracting the activity increment of a component $i$ in the shadow model from that of its real counterpart, we get the normalized (or adaptive) activity increment $\Delta^N_i(t)$:

\begin{align}
    \Delta_i^R (t) =
    \left\{
        \begin{array}{ll}
            1 & \text{if component } i \text{ exists in the real model at } t \\
            0 & \text{otherwise}
        \end{array}
    \right.
    \label{eq-delta-r}
\end{align}

\begin{align}
    \Delta_i^S (t) =
    \left\{
        \begin{array}{ll}
            1 & \text{if component } i \text{ exists in the shadow model at } t \\
            0 & \text{otherwise}
        \end{array}
    \right.
    \label{eq-delta-s}
\end{align}

\begin{align}
    \Delta_i^N (t) = \Delta_i^R (t) - \Delta_i^S (t)
    \label{eq-delta-n}
\end{align}

We can use these normalized increments to calculate the normalized cumulative evolutionary activity of each component $i$:

\begin{align}
    a_i^N (t) =
    \left\{
        \begin{array}{ll}
            \sum_{\tau = 0}^{t} \Delta_i^N (\tau) & \text{if component } i \text{ exists in the real model at } t \\
            0 & \text{otherwise}
        \end{array}
    \right.
    \label{eq-a-n}
\end{align}

From there, one can compute the total and median normalized cumulative evolutionary activity:

\begin{align}
    A_{cum}^N (t) = \sum_{i\text{: component $i$ exists in the real model at $t$}} a_i^N (t)
    \label{eq-A-n_cum}
\end{align}

\begin{align}
    \tilde{A}_{cum}^N (t) = \operatorname*{Median}\limits_{i \text{: component $i$ exists in the real model at $t$}} a_i^N (t)
    \label{eq-A-n_cum_median}
\end{align}

Using the normalized activities, we can also compute the new activity:

\begin{align}
    A_{new}^N (t) = \frac{1}{D^R (t)} \sum_{i \text{: component $i$ is ``newly adaptively-significant''}} a_i^N (t)
    \label{eq-A-n_new}
\end{align}

where an adaptively-significant component is defined as a component whose activity is greater than the absolute value of the most negative adaptive activity, and where $D^R$ is the component diversity, i.e. the number of components currently present in the real model: 

\begin{align}
    D^R (t) = | \{ i : a_i (t) > 0 \} |
    \label{eq-diversity-r}
\end{align}

Step 3 of Channon's procedure involves using the statistics of step 2 to do another open-endedness test. The test succeeds if and only if the total and median normalized cumulative evolutionary activities ($A_{cum}^N (t)$ and $\tilde{A}_{cum}^N (t)$) are unbounded and if the new evolutionary activity $A_{new}^N (t)$ is persistently positive.

\subsection{Implementation}

We ran the simulation 20 times for 2 million time steps each. Each run was used to compute evolutionary activity statistics for both step 1 and step 2. 

The program only computes activity increments of evolved agents, that is, it does not take into account activity increments of first-generation agents. This is because first-generation agents are not the result of evolution and are therefore not relevant for evaluating the open-endedness of the simulator.

\section{Results}
\label{section/results}

Here we present the evolutionary statistics of ToLSim over the 20 runs\footnote{The raw dataset is available at \url{https://figshare.com/articles/dataset/Data_including_evolutionary_activity_of_20_runs_of_the_Tree_of_Life_Simulation_/31443793?file=62289298} or using the DOI \texttt{10.6084/m9.figshare.31443793}.}.
We classify the total and median statistics as ``bounded'' when the curve seems to stagnate starting from some instant, and as ``unbounded'' when it appears to grow rather constantly.
We also classify some results as ``unclear'', i.e. not showing a clear constant growth or a clear stagnation.
Similarly, we classify the \textit{new activity} statistic as ``persistently positive'', ``not persistently positive'' or ``unclear'' according to the curve profiles.
The \hyperref[appendix]{Appendix} shows the complete results.

\subsection{Step 1: basic evolutionary activity statistics}

\begin{figure}[!htbp]
    \begin{subfigure}{0.3\textwidth}
    \centering
    \includegraphics[width=0.9\linewidth]{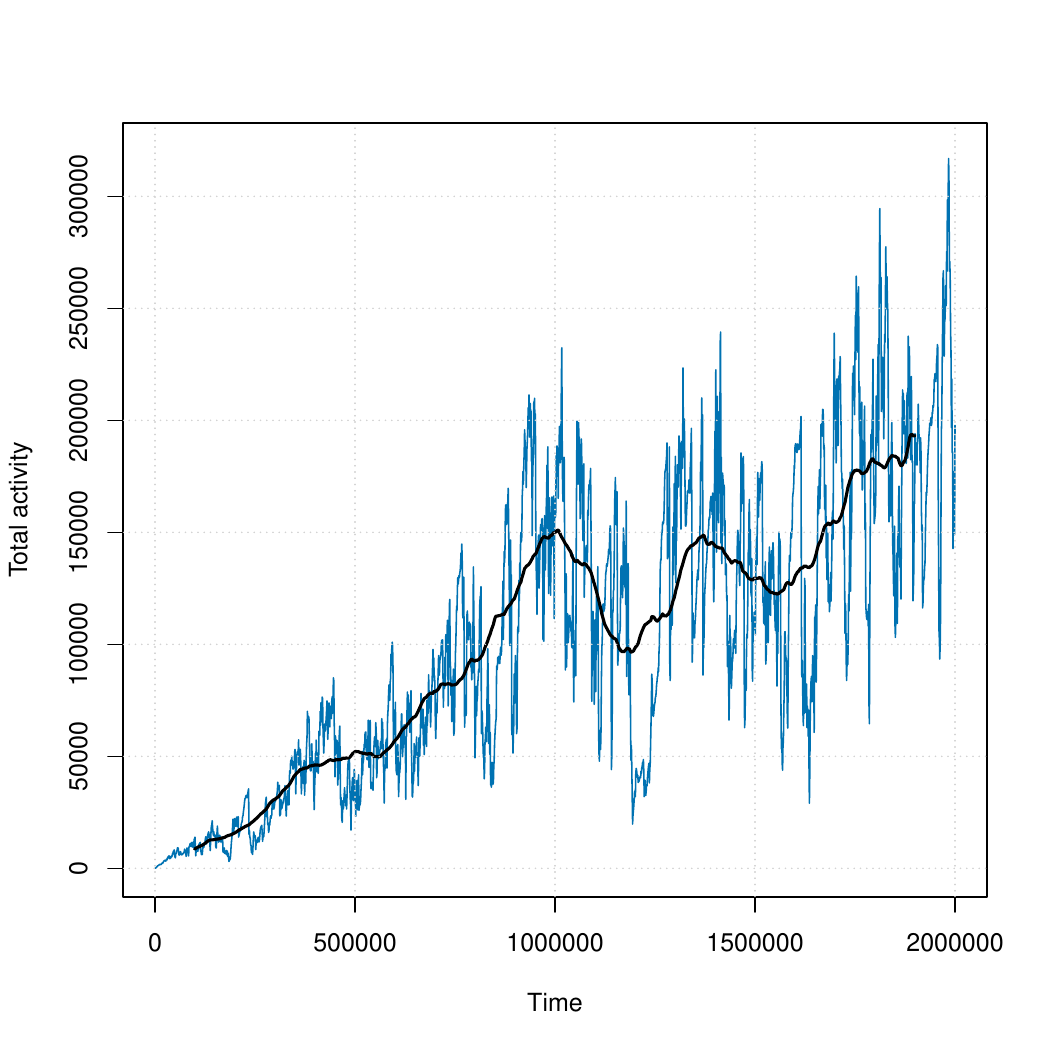}
    \caption{unbounded}
    \label{fig/step-1-unbounded}
    \end{subfigure}
    \begin{subfigure}{0.3\textwidth}
    \centering
    \includegraphics[width=0.9\linewidth]{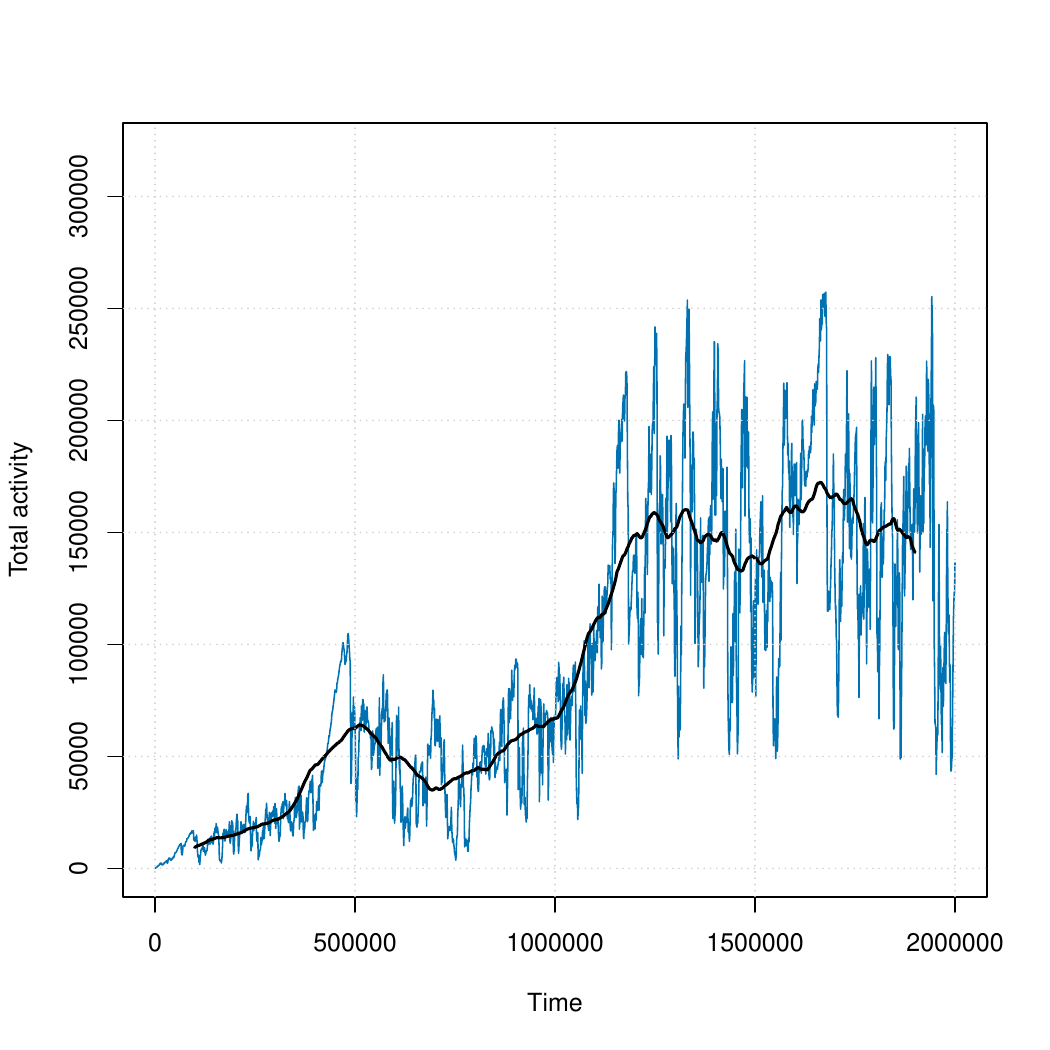}
    \caption{bounded}
    \label{fig/step-1-bounded}
    \end{subfigure}
    \begin{subfigure}{0.3\textwidth}
    \centering
    \includegraphics[width=0.9\linewidth]{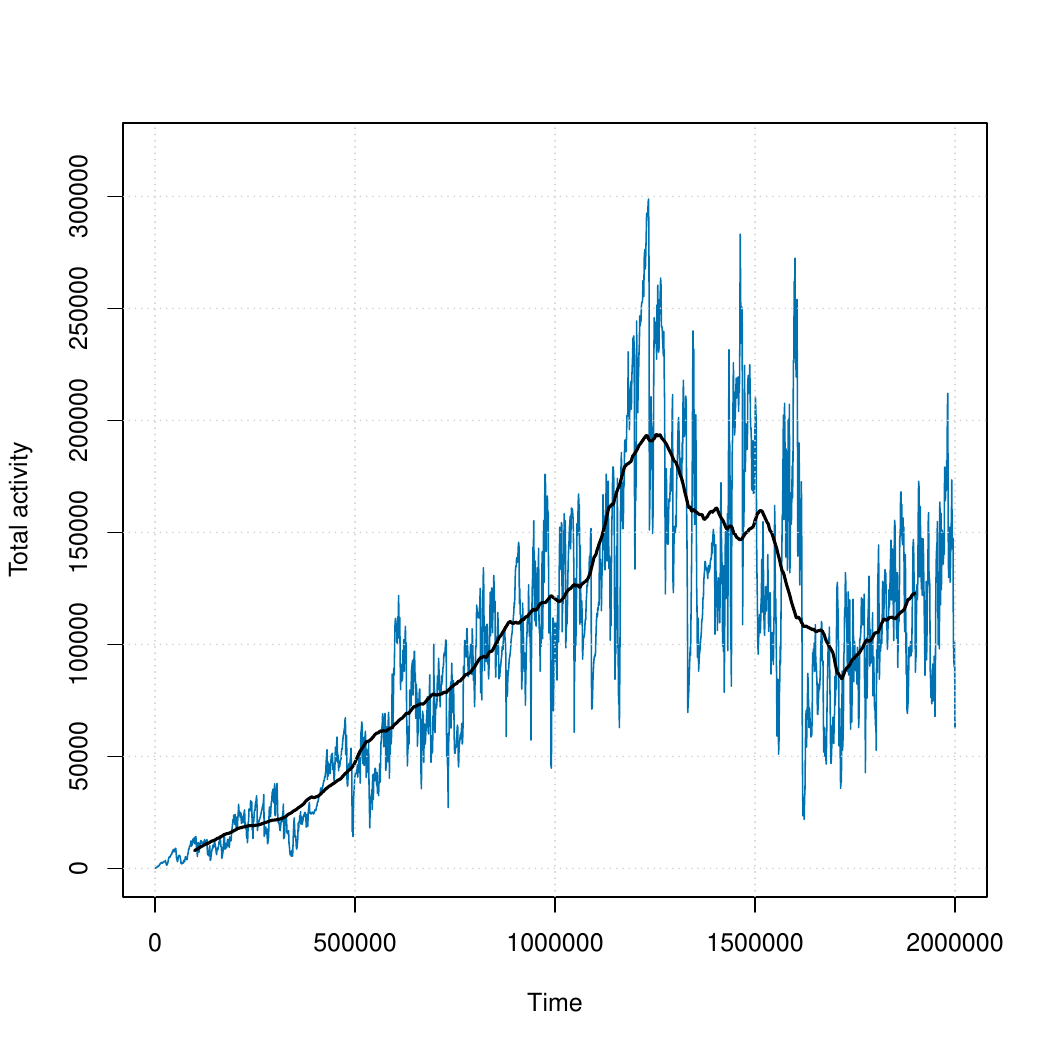}
    \caption{unclear}
    \label{fig/step-1-unclear}
    \end{subfigure}
    \caption{Total cumulative activity curves of runs n°1, n°12 and n°20, each illustrating a typical profile found across the 20 runs.}
    \label{fig/step-1}
\end{figure}

Total cumulative evolutionary activity seems unbounded in 8 runs (fig. \ref{fig/step-1-unbounded}), bounded in 4 runs (fig. \ref{fig/step-1-bounded}), and unclear in 8 runs (fig. \ref{fig/step-1-unclear}). Overall, this statistic appears unbounded in 40\% of the runs. ToLSim therefore can be said to partially pass the test of step 1.

\subsection{Steps 2 and 3: component-normalized evolutionary activity statistics}

In all 20 runs, the \textit{new activity} does not appear to be constantly positive; in fact, each run shows a mostly null \textit{new activity}. This single result is sufficient to conclude that the test of step 3 is not passed by ToLSim. 

\begin{figure}[!htbp]
    \begin{subfigure}{0.3\textwidth}
    \centering
    \includegraphics[width=0.9\linewidth]{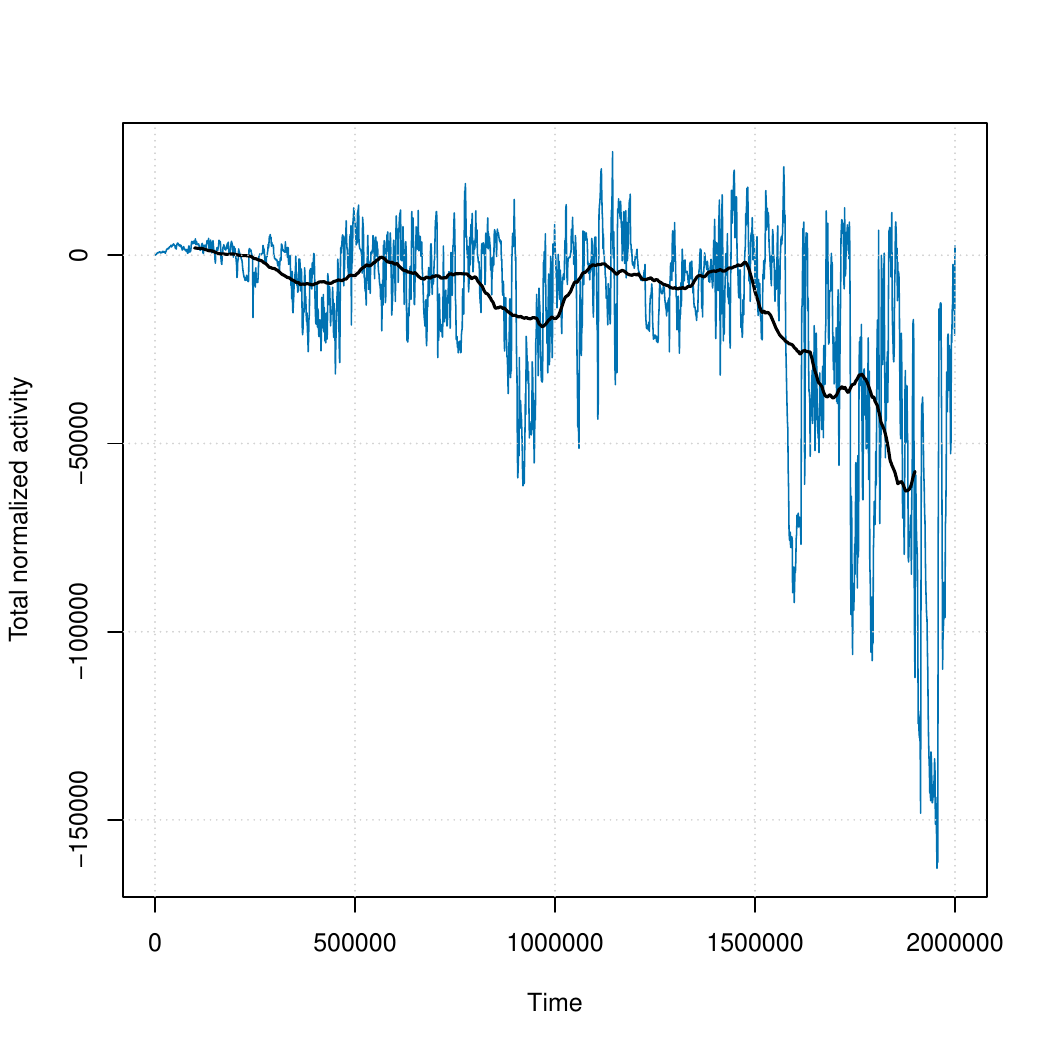}
    \caption{total normalized activity}
    \label{fig/step-2-original-formula-total-normalized-activity}
    \end{subfigure}
    \begin{subfigure}{0.3\textwidth}
    \centering
    \includegraphics[width=0.9\linewidth]{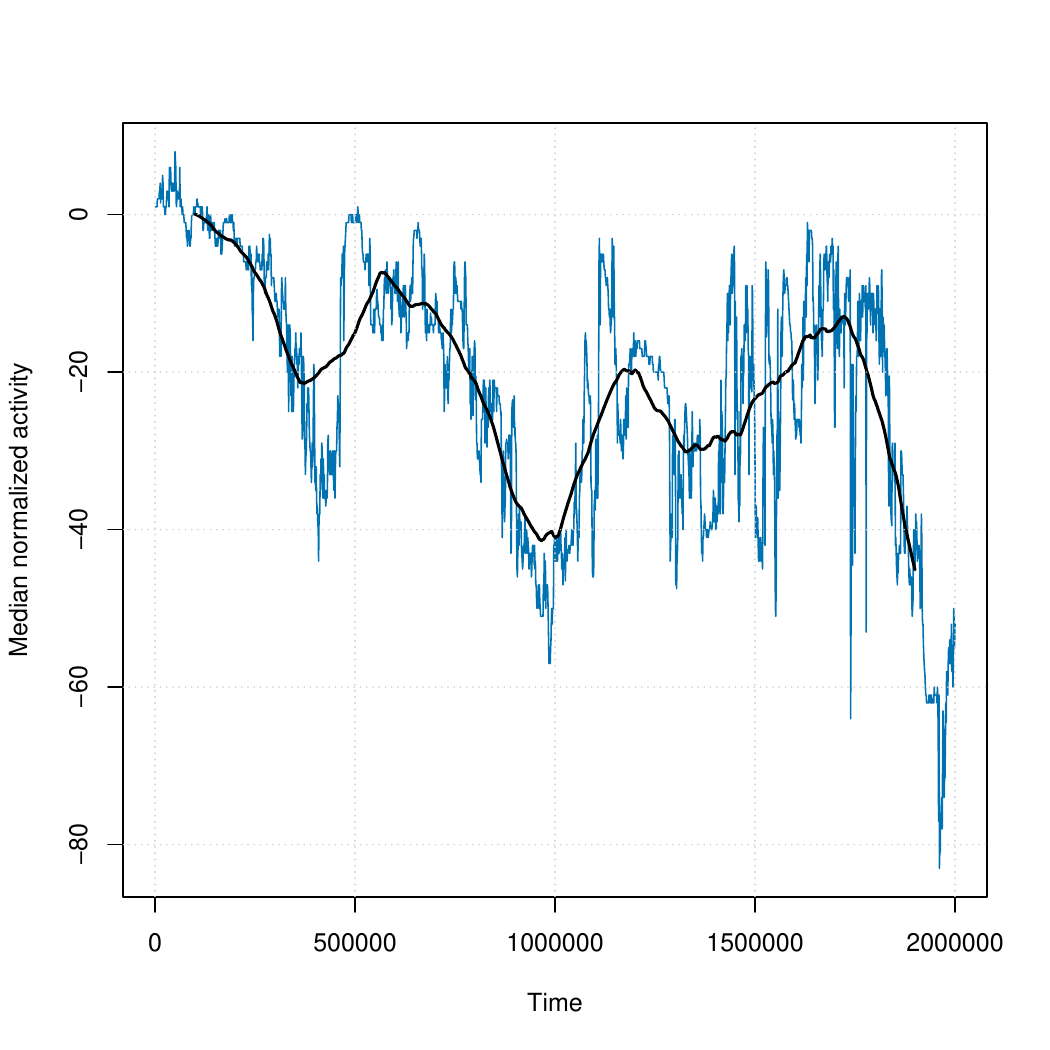}
    \caption{median normalized activity}
    \label{fig/step-2-original-formula-median-normalized-activity}
    \end{subfigure}
    \begin{subfigure}{0.3\textwidth}
    \centering
    \includegraphics[width=0.9\linewidth]{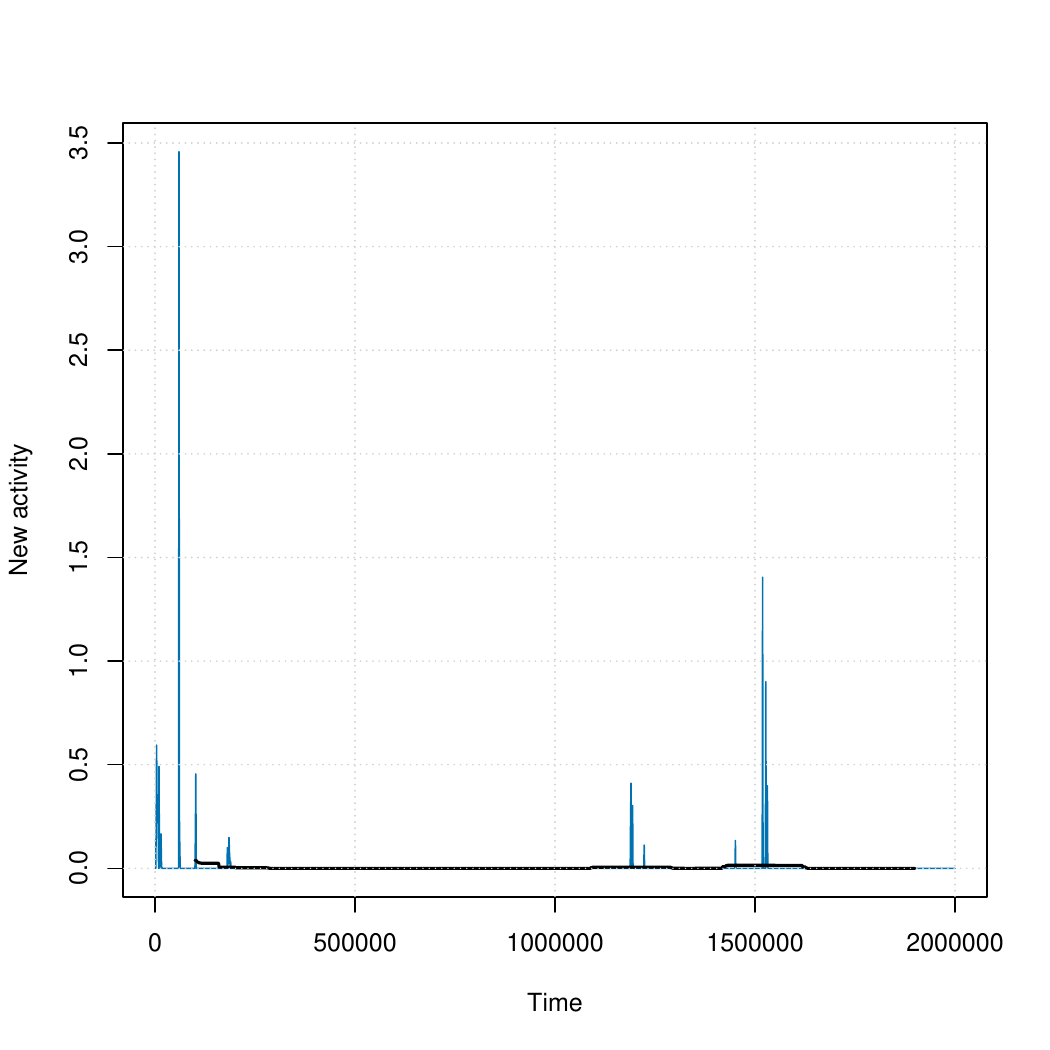}
    \caption{new activity}
    \label{fig/step-2-new-activity}
    \end{subfigure}
    \caption{Total normalized cumulative activity, median normalized cumulative activity and new activity of run n°1. This run is representative of all runs.}
    \label{fig/step-2-original-formula}
\end{figure}

All runs of ToLSim show bounded (and even decreasing) total and median normalized cumulative activities. Figure \ref{fig/step-2-original-formula} shows the graphs of the first run as a typical illustration of the trends of the three statistics. We can see that new activity stays null and that total and median normalized activities tend to decrease, going in the negative numbers. The reason why these latter statistics are negative may be that most components are only present in the shadow model, since selection is random there and allows for more diversity.

\section{Conclusion}

We conducted an experiment on the Tree of Life Simulation (ToLSim) artificial life software, testing for open-endedness at the level of genes. We applied steps 1, 2 and 3 of Channon's procedure for testing open-endedness \citep{channon-2024-procedure}, looking at statistical trends related to complexity and novelty. We found that 8 out of 20 runs passed the first test (step 1); none passed the second test (step 3).

The failure to pass the second test may be due to ToLSim not being open-ended. It could also be because we analyzed the program at the level of genes. Indeed, ToLSim might pass the second test at the level of individuals or the level of species. There is also the possibility that the settings of the simulation are not optimal for open-endedness. Settings that could be changed include the size of the map, the energy input or the cost of being alive.

This experiment could be improved.
Indeed, defining components in advance has limitations. Indeed, it prevents identifying and analyzing new components on the fly, therefore limiting open-endedness tests. A solution discussed by \citep{taylor-2016-open} would be to define models at the scale of physics and chemistry, rather than on the scale of genes or individuals.
Another limitation is that the statistics have been classified based on eye sight. A more systematic method would be beneficial.

Managing to create an open-ended artificial life software could have huge benefits.
It could not only allow us to have a better understanding of life and open-endedness, but also to create algorithms that can generate novel ideas in various areas of life, such as art, video game, or science.
It is unclear whether programs as open-ended as Earth's biosphere are possible, but the present paper, as well as other findings such as those regarding Geb, can spark hope.

\section{Acknowledgements}

We would like to thank Alastair Channon for thorough discussions and advice while during the data analysis. Many thanks to Hyoyeon Lee and Jane Adams, who contributed useful visualizations to the ToLSim project. 

\section*{Appendix}
\label{appendix}

\begin{center}
\rotatebox{90}{%
\begin{minipage}{\textheight} 
\centering
\small
\renewcommand{\arraystretch}{1.5} 
\begin{tabular}{c|p{2.5cm}|p{3cm}|p{3cm}|p{3.5cm}|p{3.5cm}|p{1.5cm}}
\hline
Run &
$A_{cum} (t)$ &
$A_{cum}^N (t)$ &
$\tilde{A}_{cum}^N (t)$ &
$A_{new}^N (t)$ \\
\hline
1  & Unbounded & Bounded & Bounded & Null \\
2  & Unclear   & Bounded & Bounded & Null \\
3  & Unbounded & Bounded & Bounded & Null \\
4  & Bounded   & Bounded & Bounded & Null \\
5  & Bounded   & Bounded & Bounded & Null \\
6  & Unbounded & Bounded & Bounded & Null \\
7  & Unbounded & Bounded & Bounded & Null \\
8  & Unclear   & Bounded & Bounded & Null \\
9  & Bounded   & Bounded & Bounded & Null \\
10 & Unbounded & Bounded & Bounded & Null \\
11 & Unclear   & Bounded & Bounded & Null \\
12 & Bounded   & Bounded & Bounded & Null \\
13 & Unbounded & Bounded & Bounded & Null \\
14 & Unbounded & Bounded & Bounded & Null \\
15 & Unclear   & Bounded & Bounded & Null \\
16 & Unclear   & Bounded & Bounded & Null \\
17 & Unbounded & Bounded & Bounded & Null \\
18 & Unclear   & Bounded & Bounded & Null \\
19 & Unclear   & Bounded & Bounded & Null \\
20 & Unclear   & Bounded & Bounded & Null \\
\hline
\end{tabular}

\vspace{0.5em}
\captionof{table}{Trends of the evolutionary statistics of the Tree of Life Simulator (ToLSim).}
\label{table/trends}
\end{minipage}%
}
\end{center}

\clearpage 
\bibliography{refs}

\end{document}